\pdfoutput=1
\documentclass[fleqn,usenatbib,usedcolumn]{mnras}
\usepackage[british]{babel}             
\hypersetup{pdfauthor={D. A. Green},
            pdftitle={Constraints on the distribution of supernova 
                      remnants with Galactocentric radius},
            pdfkeywords={supernova remnants, Galaxy: structure, ISM: structure},
            bookmarksnumbered=true}
\volume{{\rm in press}}
\usepackage{txfonts}
\makeatletter
\re@DeclareMathSymbol{\Sigma}{\mathalpha}{letters}{6}
\makeatother
\let\la=\lesssim 
\usepackage[T1]{fontenc}                
\usepackage{graphicx}                   
\usepackage{booktabs}                   
\def\SNR(#1.#2)#3(#4.#5){{G#1${\cdot}$#2$#3$#4${\cdot}$#5}}
\newcommand{\SigmaUnit}{{\rm W~m^{-2}\,Hz^{-1}\,sr^{-1}}}
\title[Constraints on the distribution of Galactic SNRs]{Constraints on
the distribution of supernova remnants with Galactocentric radius}

\author[Green]{D.~A.~Green\thanks{email: {\tt dag@mrao.cam.ac.uk}}\\
 Astrophysics Group, Cavendish Laboratory, 19 J.~J.~Thomson Avenue,
 Cambridge CB3 0HE}

\date{Accepted ---; received ---; in original form ---}

\pubyear{2015}
\begin{document}

\label{firstpage}

\pagerange{\pageref{firstpage}--\pageref{lastpage}}

\maketitle

\begin{abstract}
Supernova remnants (SNRs) in the Galaxy are an important source of energy
injection into the interstellar medium, and also of cosmic rays. Currently
there are 294 known SNRs in the Galaxy, and their distribution with
Galactocentric radius is of interest for various studies. Here I discuss some
of the statistics of Galactic SNRs, including the observational selection
effects that apply, and difficulties in obtaining distances for individual
remnants from the `$\Sigma{-}D$' relation. Comparison of the observed Galactic
longitude distribution of a sample of bright Galactic SNRs -- which are not
strongly affected by selection effects -- with those expected from models is
used to constrain the Galactic distribution of SNRs. The best-fitting
power-law/exponential model is more concentrated towards the Galactic centre
than the widely used distribution obtained by \citet{1998ApJ...504..761C}.
\end{abstract}

\begin{keywords}
  supernova remnants -- Galaxy: structure -- ISM: structure
\end{keywords}

\section{Introduction}\label{s:introduction}

There are currently 294 supernova remnants (SNRs) known in the Galaxy
\citep{2014BASI...42...47G}. These are an important source of the
injection of energy and heavy elements into the interstellar medium, and
are believed to be the site of acceleration of cosmic rays, at least up
to $10^{15}$~eV (e.g.\ \citealt{2014BrJPh..44..415B}). Consequently the
distribution of SNRs in the Galaxy is of interest for a variety of
studies of the Galaxy (e.g.\ \citealt{2011APh....35..211L,
2012ApJ...752...68V, 2014ApJ...785..129K, 2015JCAP...03..038C}).

It is not straightforward to construct a Galactic distribution directly
from properties of catalogued SNRs. This is not only because distances
are not available for most Galactic SNRs, but also because of the
observational selection effects that apply to the current catalogue of
SNRs. It is expected \citep{2011MNRAS.412.1473L} that most SNRs are from
massive stars (i.e.\ types II/Ib/Ic), which also produce pulsars, and
hence the distribution of SNRs will be closely related to that of star
forming regions or pulsars.

Some statistics of the current Galactic SNR catalogue, the selection
effects that apply to their identification, and issues related to the
derivation of distances to individual SNRs from the `$\Sigma{-}D$'
relation are discussed in Section~\ref{s:background}.
Section~\ref{s:radial} presents constraints on the Galactic distribution
of SNRs from comparison of the $l$-distribution of bright SNRs with
various models. This includes comparison with the distribution from
\citeauthor{1998ApJ...504..761C} (1998), hereafter CB98 (see also
\citealt{1996A&AS..120C.437C}), which has been widely used. The
conclusions are summarised in Section~\ref{s:conclusions}. Preliminary
results have been presented in \citet{2012AIPC.1505....5G,
2014IAUS..296..188G}, but here a more detailed analysis is made.

\begin{figure*}
\centerline{\includegraphics[width=15.5cm]{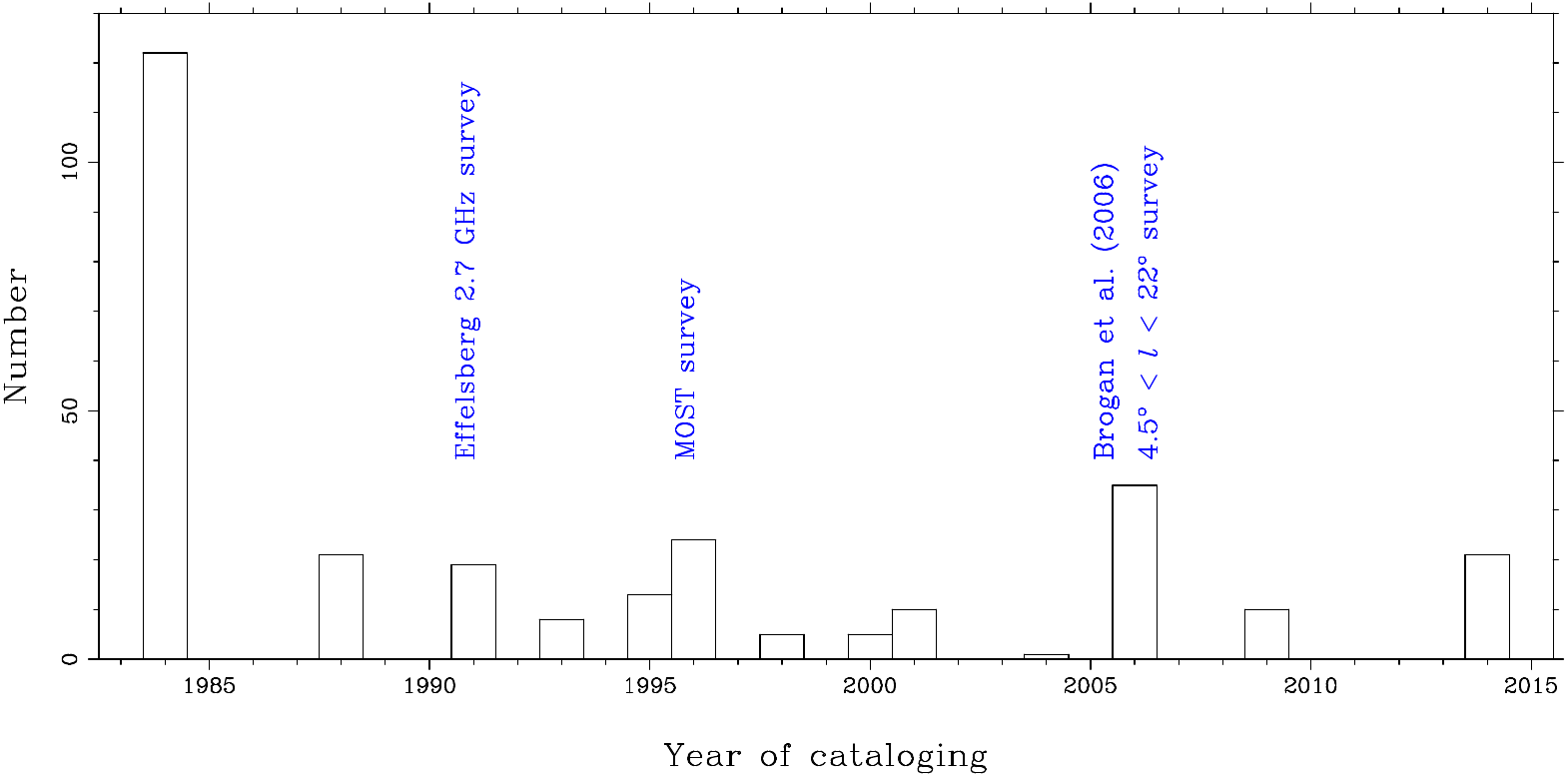}}
\caption{Histogram of the dates when Galactic SNRs were first
catalogued.\label{f:h-year}}
\end{figure*}

\section{Galactic SNRs}\label{s:background}

\subsection{The catalogue of SNRs}\label{s:catalogue}

I have produced several catalogues of Galactic SNRs, with published
versions in \citet{1984MNRAS.209..449G, 1988Ap&SS.148....3G,
1991PASP..103..209G, 1996ssr..conf..419G, 2004BASI...32..335G,
2009BASI...37...45G, 2014BASI...42...47G}, and also various versions
online since 1993\footnote{see:
\texttt{http://www.mrao.cam.ac.uk/surveys/snrs/}}. Of the 294 remnants
in the most recent version of the catalogue, most have been first
identified at radio wavelengths, or -- if identified at other wavebands
-- have subsequently been detected at radio wavelengths. But there are
22 remnants in the catalogue that have not yet been detected at radio
wavelengths, or not sufficiently well observed to provide an integrated
radio flux density. At optical and X-ray wavelengths only about 40\% and
30\%, respectively, of the catalogued SNRs have been detected (which is
not surprising, due to Galactic absorption that affects these
wavelengths). Thus, it is selection effects at radio wavelengths that
are dominant when considering the completeness of the catalogue.
Figure~\ref{f:h-year} shows a histogram of the date of inclusion of a
remnant in the catalogue (the larger number of entries in the first
version of the catalogue, from 1984, were largely taken from earlier
Galactic SNR catalogues). This shows that major radio surveys have been
the cause of significant increases in the identification of Galactic
SNRs, notably:
\begin{enumerate}
\item the Effelsberg 2.7-GHz survey \citep{1990A&AS...85..633R,
1990A&AS...85..691F}, covering $-2.6^\circ < l < 240^\circ$, $|b| <
5^\circ$, with a resolution of $\approx 4.3$~arcmin;
\item the MOST 843-GHz survey \citep{1999ApJS..122..207G}, covering
$245^\circ < l < 255^\circ$, $|b| < 1\fdg5$, with a resolution of
$\approx 0.7$~arcmin at best;
\item \citet{2006ApJ...639L..25B}'s survey of a small region of the 1st
Galactic quadrant -- $4\fdg5 < l < 22^\circ, |b| < 1\fdg25$ -- at
multiple radio wavelengths, with a resolution of $\approx 0.6$~arcmin at
327~MHz, which also used infrared observations to help discriminate
between different types of sources.
\end{enumerate}
In the latest revision of the Galactic SNR catalogue, 21 new remnants
were added to the catalogue. Of these 13 have an integrated radio flux
density at 1~GHz in the catalogue, and hence a surface brightness,
$\Sigma$ can be
calculated; all of these 13 remnants are faint, with $\Sigma_{\rm 1~GHz}
< 8 \times 10^{-21}~\SigmaUnit$. The other 8 newly catalogued remnants
do not yet have radio flux densities reported in the literature, as they
were identified at optical or X-ray wavelengths (e.g.\ \SNR(38.7)-(1.3),
\SNR(65.8)-(0.5), \SNR(66.0)-(0.0), \SNR(67.6)-(0.0), \SNR(67.6)+(0.9)
and \SNR(67.8)+(0.5) were identified from optical observations by
\citealt{2013MNRAS.431..279S} -- for these SNRs integrated radio flux
densities have not been yet published).

\begin{figure}
\centerline{\includegraphics[width=8.5cm]{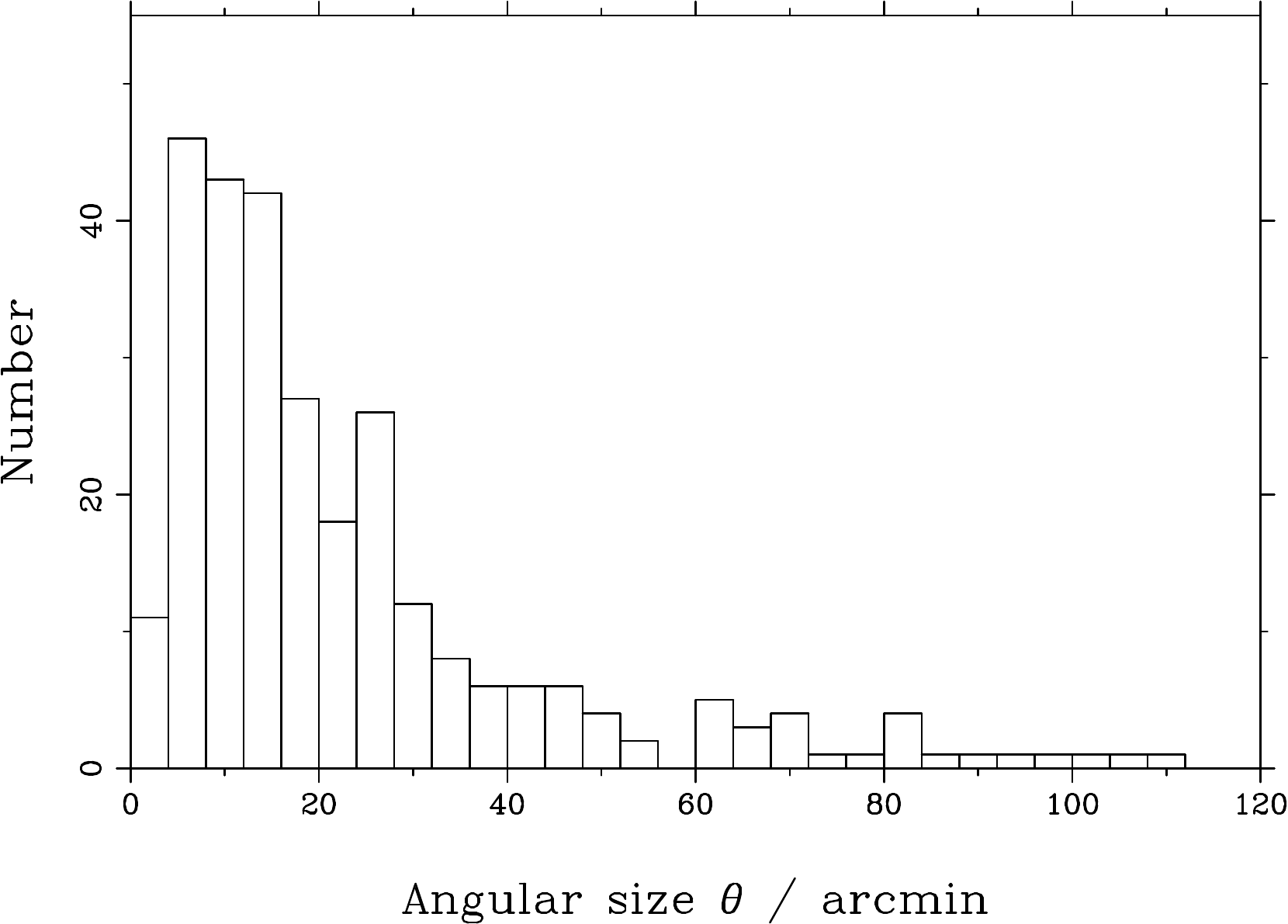}}
\caption{Histogram of the angular size, $\theta$, of the catalogued
Galactic SNRs (there are also 10 large remnants with angular sizes
greater than $2^\circ$ which are not included in this
plot).}\label{f:h-size}
\end{figure}

In order to derive directly the distribution of SNRs in the Galaxy both:
(i) correction for the incompleteness of the current catalogue of SNRs
due to observational selection effects, and (ii) knowledge of the
distances to each known SNR is required. I discuss below the selection
effects that apply, at radio wavelengths, to the current SNR catalogue,
and also some issues with the derivation of distances to individual SNRs
using the `$\Sigma{-}D$' relation.

\subsection{Selection effects}\label{s:selection}

As noted above, most Galactic SNRs are identified at radio wavelengths,
and the selection effects that apply -- as discussed previously, e.g.\
\citet{1991PASP..103..209G, 2005MmSAI..76..534G} -- are (i)
intrinsically faint remnants (i.e.\ low surface brightness) are
difficult to identify, and (ii) physically small but distant remnants
are difficult to recognise as SNRs, due to their small angular sizes.

I have previously argued \citep{2004BASI...32..335G} for a nominal
surface brightness completeness limit of $\approx 10^{-20}~\SigmaUnit$.
(which is equivalent to $\approx 65$~mJy per 1-arcmin circular beam).
However, a higher limit probably applies close to the Galactic Centre
(GC), where the background Galactic radio emission is brighter than
elsewhere. The validity of this value for the approximate surface
brightness completeness of the current SNR catalogue is discussed
further below.

In addition to the difficulty of identifying low-surface brightness
SNRs, it is also difficult to identify small angular size remnants. For
example, the Effelsberg 2.7-GHz survey has a resolution of 4.3~arcmin,
and SNRs would have to be several times this angular size for their
structure to be recognised. \citet{1988srim.conf..293R} reported 32 new
supernova remnants in the first part of the area covered by the
Effelsberg 2.7-GHz survey ($-2\fdg6 \le l \le 76^\circ$, $|b| \le
5^\circ$, subsequently published in \citealt{1990A&AS...85..633R}). Of
these 30 are resolved, with angular diameters of 16~arcmin or more,
i.e.\ several times the resolution of the survey. The other two sources
have small angular size, and were thought to be SNRs on the basis of
other, higher resolution targeted observations: (i) \SNR(54.1)+(0.3),
then thought to be a small, $\approx 1.5$~arcmin `filled centre' remnant
(see \citealt{1985MNRAS.216..691G, 1985A&A...151L..10R}), but is now
catalogued as a slightly larger $\approx 12$~arcmin composite remnant,
since a faint X-ray halo was detected by \citet{2010A&A...520A..71B},
and (ii) \SNR(70.7)+(1.2), which was reported as a SNR by
\citet{1985A&A...151L..10R}, but this identification was not supported
by subsequent observations (e.g.\ \citealt{1986MNRAS.219P..39G,
1995ApJ...449L.127O, 2007ApJ...665L.135C} and references therein). Given
that SNRs are expected to have a continuous range of physical diameters,
and are seen at a range of distances in the Galaxy, then a smooth
distribution of angular sizes is expected. Figure~\ref{f:h-size} shows a
histogram of the angular size, $\theta$, of catalogued SNRs, which
clearly shows a sharp decrease at small angular sizes ($<4$~arcmin),
due to the difficulty in identifying small SNRs.

\begin{figure}
\leftline{\quad\large a)}
\vskip6pt
\centerline{\includegraphics[width=8.5cm]{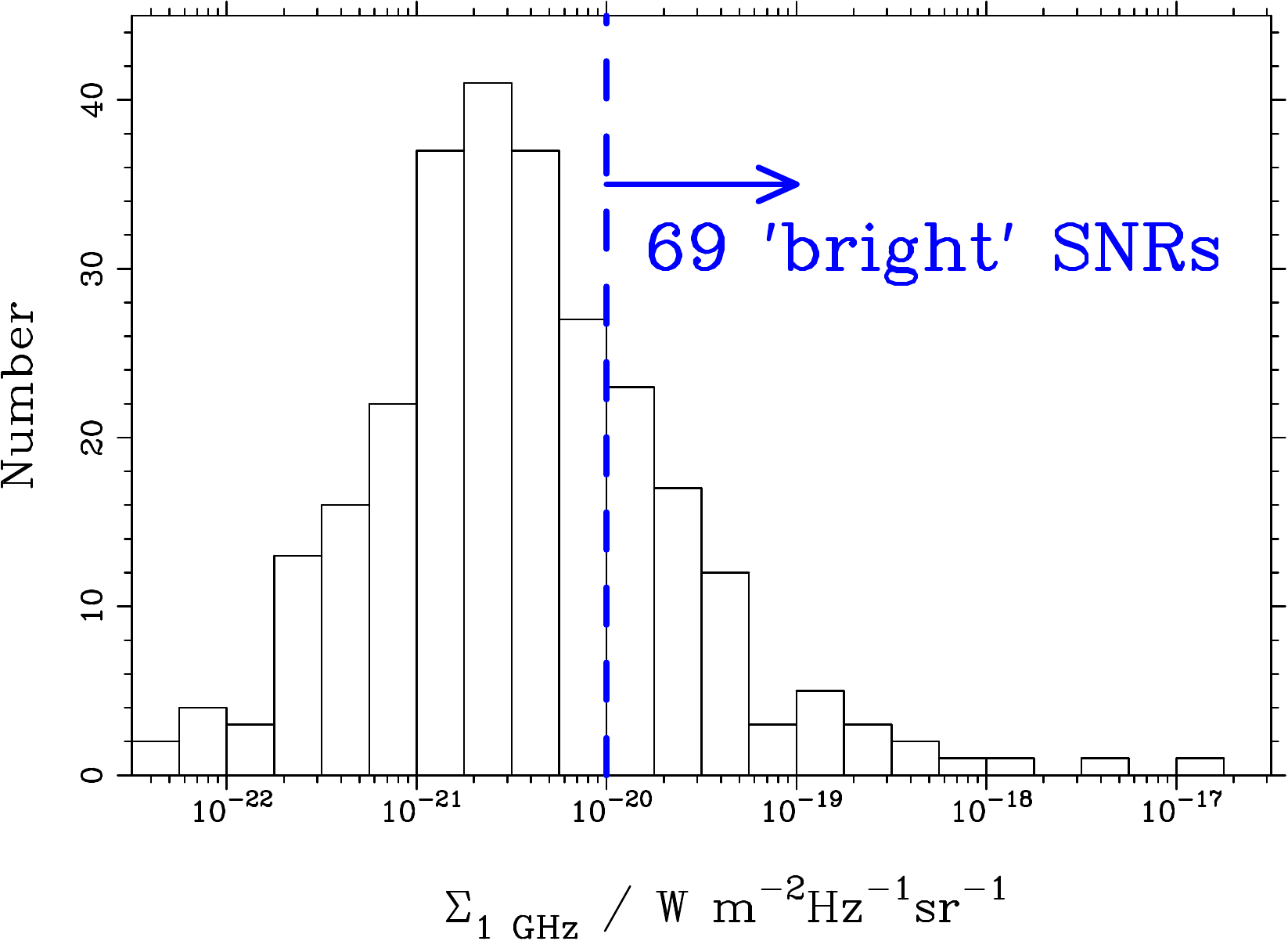}}
\vskip6pt
\leftline{\quad\large b)}
\vskip6pt
\centerline{\includegraphics[width=8.5cm]{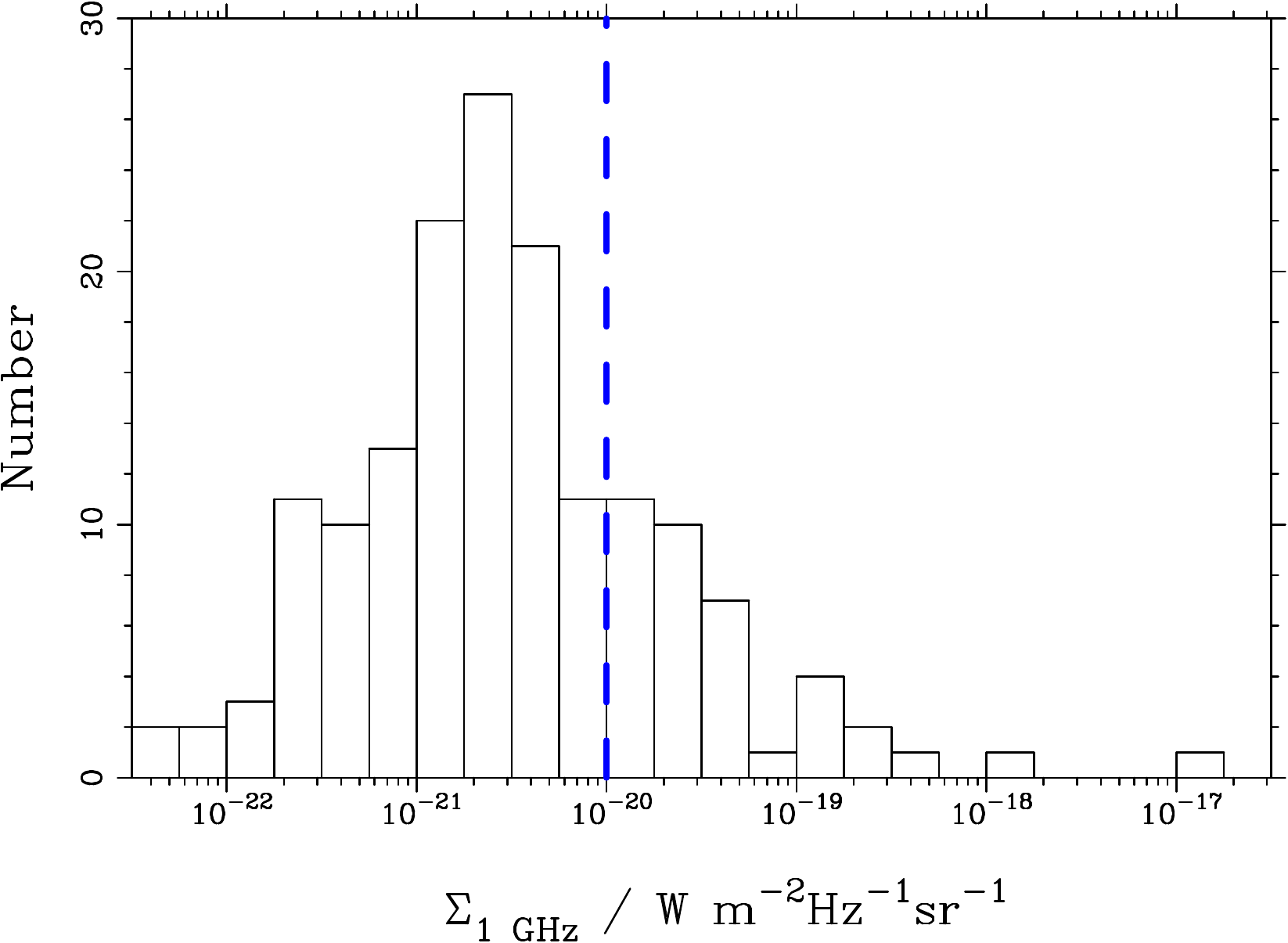}}
\vskip6pt
\leftline{\quad\large c)}
\vskip6pt
\centerline{\includegraphics[width=8.5cm]{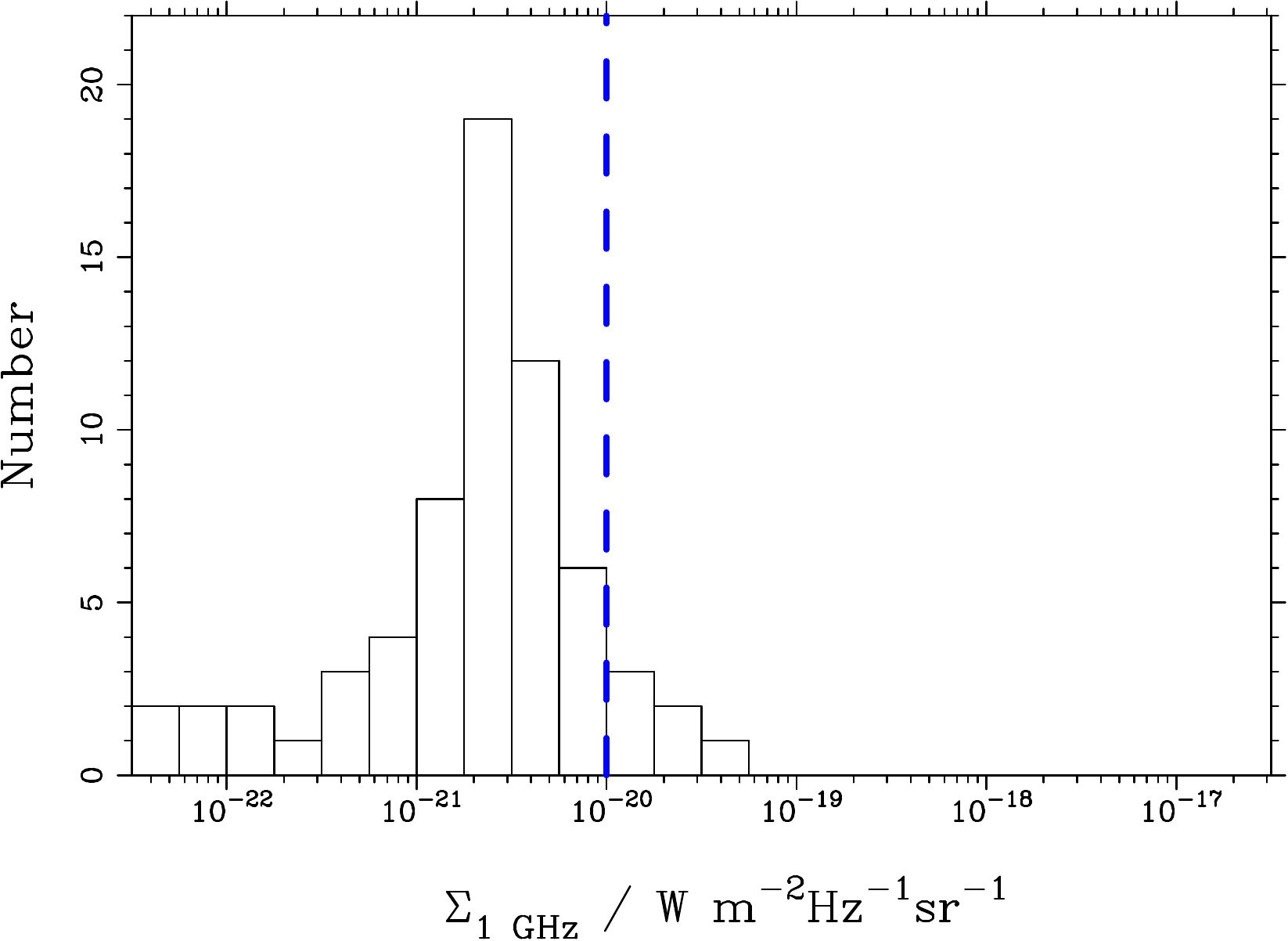}}
\caption{Histograms of surface brightness at 1~GHz for catalogued SNRs with a
radio flux density at 1~GHz for: (a) SNRs; (b) SNRs in the region covered by
the Effelsberg 2.7-GHz survey (i.e.\ $-2^\circ < 240^\circ$, $|b| < 5^\circ$),
(c) SNRs in the Effelsberg survey region identified since
1992.\label{f:h-sigma}}
\end{figure}

\begin{table}
\caption{SNRs in Effelsberg 2.7~GHz survey region with $\Sigma_{\rm
1~GHz} > 10^{-20}$ $\SigmaUnit$ identified since
1992.\label{t:sigma-post92}}
\smallskip
\centering
\def\chead#1{\multicolumn{1}{c}{#1}}
\begin{tabular}{cc.c}\toprule
  name  & $\theta$ & \chead{$S_{\rm 1~GHz}$} & $\log\Sigma_{\rm 1~GHz}$ \\
        & / arcmin & \chead{/ Jy}            & / $\SigmaUnit$           \\\midrule
  \SNR(0.3)+(0.0) &  $15\times8$  &  22   & $-19.56$ \\             
  \SNR(1.0)-(0.1) &       8       &  15   & $-19.45$ \\             
  \SNR(6.5)-(0.4) &      18       &  27   & $-19.90$ \\             
 \SNR(12.8)-(0.0) &       3       &   0.8 & $-19.87$ \\             
 \SNR(18.1)-(0.1) &       8       &   4.6 & $-19.97$ \\             
 \SNR(20.4)+(0.1) &       8       &   9.0 & $-19.67$ \\\bottomrule  
\end{tabular}
\end{table}

Although there are some small angular size remnants in the catalogue
identified from high resolution observations -- notably \SNR(1.9)+(0.3) (e.g.\
\citealt{2008MNRAS.387L..54G, 2008ApJ...680L..41R, 2014ApJ...790L..18B}), with
an angular diameter of only 1.5~arcmin, many are missing. These missing young
but distant SNRs will be on the far side of the Galaxy, and hence concentrated
towards $l = 0^\circ$, $b = 0^\circ$, where confusion due to other Galactic
sources along the line of sight makes their identification difficult.

To justify the nominal surface brightness completeness limit of $\approx
10^{-20}~\SigmaUnit$ is appropriate, Fig.~\ref{f:h-sigma} shows histograms
of: (a) 272 of the 294 catalogued SNRs which have an integrated flux
density, and hence a surface brightness at 1~GHz; (b) 171 SNRs in
the Effelsberg 2.7~GHz survey region; (c) 38 SNRs in the Effelsberg 2.7~GHz
survey region entered into my SNR catalogue since 1992, i.e.\ those not
identified from the Effelsberg 2.7~GHz survey.

\begin{figure*}
\leftline{\qquad\qquad\large a)}
\vskip6pt
\centerline{\includegraphics[width=15.5cm]{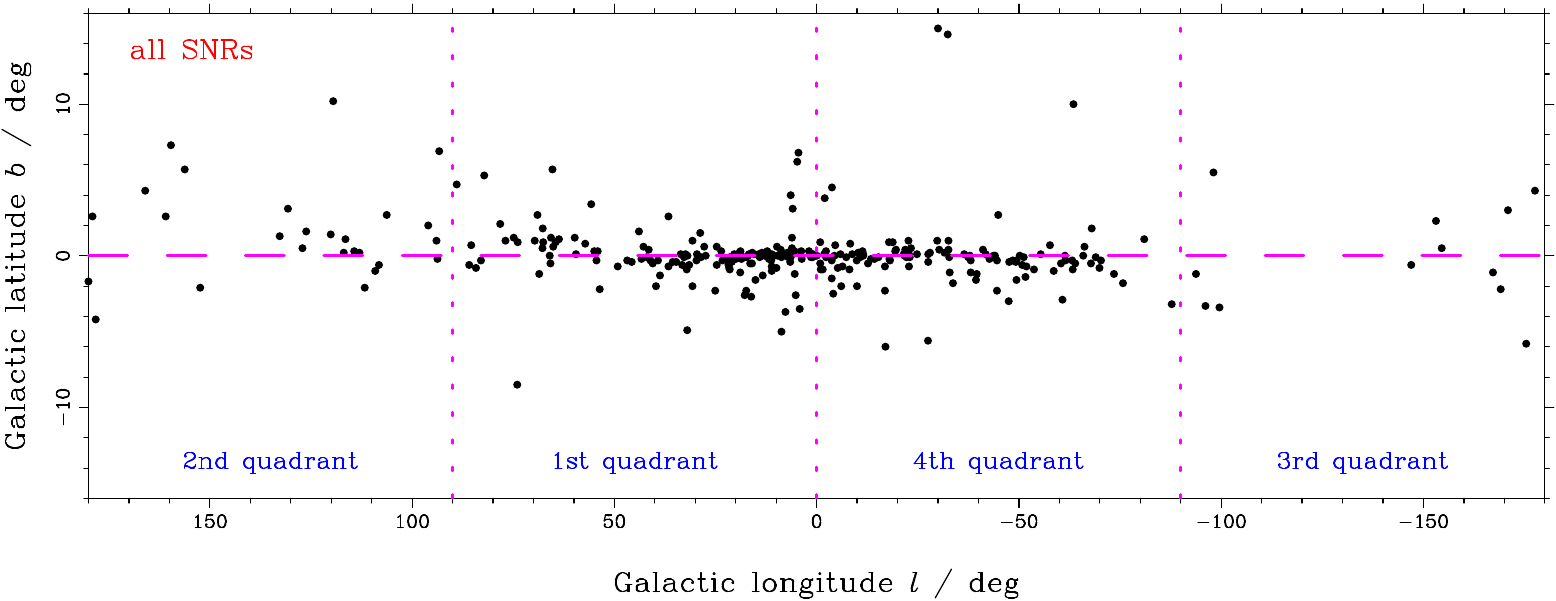}}
\vskip6pt
\leftline{\qquad\qquad\large b)}
\vskip6pt
\centerline{\includegraphics[width=15.5cm]{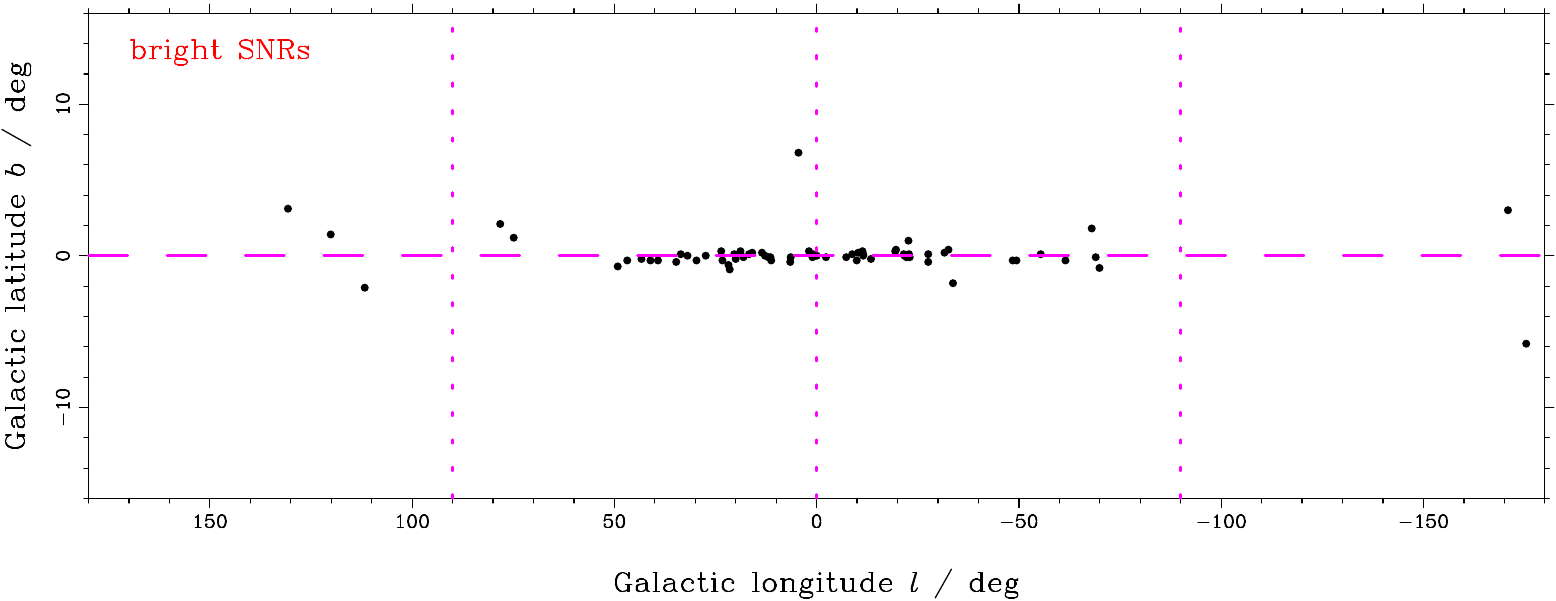}}
\caption{The distribution of SNRs in Galactic coordinates of all 294
catalogued SNRs for: (a) all SNRs, (b) the 69 `bright' SNRs with a surface
brightness above $10^{-20}~\SigmaUnit$. Note that the latitude scale is
exaggerated.\label{f:s-lb}}
\end{figure*}

\begin{figure*}
\centerline{\includegraphics[width=15.5cm]{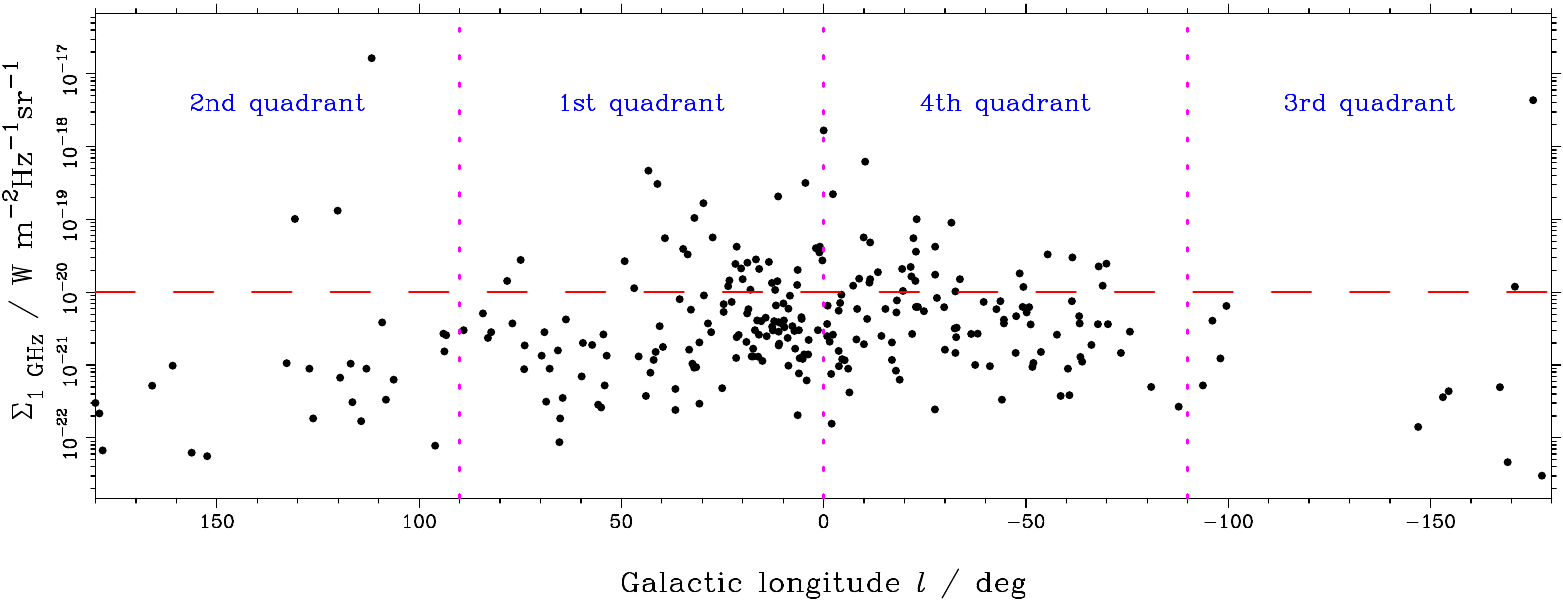}}
\caption{Radio surface brightness at 1~GHz against Galactic longitude for the
272 catalogued SNRs with radio flux densities. The nominal surface brightness
limit of $\Sigma_{\rm 1~GHz} = 10^{-20}~\SigmaUnit$ is
shown.}\label{f:s-Sigmal}
\end{figure*}

Figure~\ref{f:h-sigma}(a) shows that the majority of SNRs identified are
fainter than this nominal surface brightness limit, with 69 `bright'
remnants above the limit. The fainter remnants are more easily detected
in regions where the Galactic background emission is fainter, i.e.\ away
from $b = 0^\circ$ and $l = 0^\circ$, which is illustrated in
Figs~\ref{f:s-lb} and \ref{f:s-Sigmal}. Figure~\ref{f:s-lb} shows the
distribution in Galactic coordinates of (a) all remnants, and (b) the 69
`bright' remnants. The distribution of all remnants, without taking
the surface brightness effect into account, is very much broader in both
coordinates than that of the bright remnants (the r.m.s.\ deviations
from the Galactic Centre are $42^\circ$ and $2\fdg4$ in $l$ and $b$
respectively for all SNRs, with $36^\circ$ and $1\fdg3$ for the 69
`bright' SNRs). Figure~\ref{f:s-Sigmal} shows the surface brightness and
Galactic longitude of the 272 catalogued SNRs with radio flux densities.
In the Galactic anti-centre (i.e.\ the 2nd and 3rd Galactic quadrants)
-- where the Galactic background is low -- there is a higher proportion
of low surface brightness SNRs.

Figure~\ref{f:h-sigma}(b) and (c) show that the majority of SNRs
identified in the Effelsberg 2.7~GHz survey region from subsequent other
observations are fainter than the nominal surface brightness limit.
However, there are 6 SNRs above the surface brightness limit, which have
subsequently been identified, see Table~\ref{t:sigma-post92} (some of
these had been suggested as possible SNRs earlier). These 6 remnants are
all close to the Galactic centre, or are small angular size, so would
not have been well resolved in the Effelsberg 2.7~GHz survey, which is
necessary to be recognised as a SNR.

A surface brightness completeness limit of $\Sigma_{\rm 1~GHz} \approx
10^{-20}~\SigmaUnit$ is also consistent with observations made by
\citet{2013A&A...559A..81X}. Their observations covered $66^\circ < l <
90^\circ$, $|b| < 4^\circ$ at 5~GHz, and they used other radio and
infra-red observations to separate non-thermal and thermal components.
This multi-wavelength approach should allow fainter remnants to be
recognised than is possible using a single observation frequency. They
concluded that there were no large, i.e.\ sufficiently resolved,
remnants with $\Sigma_{\rm 1~GHz} \goa 0.37 \times 10^{-20}~\SigmaUnit$
in the region they observed.

\begin{figure}
\leftline{\qquad a)}
\centerline{\includegraphics[viewport=54 0 378 788,clip=,
              angle=270,width=8.5cm]{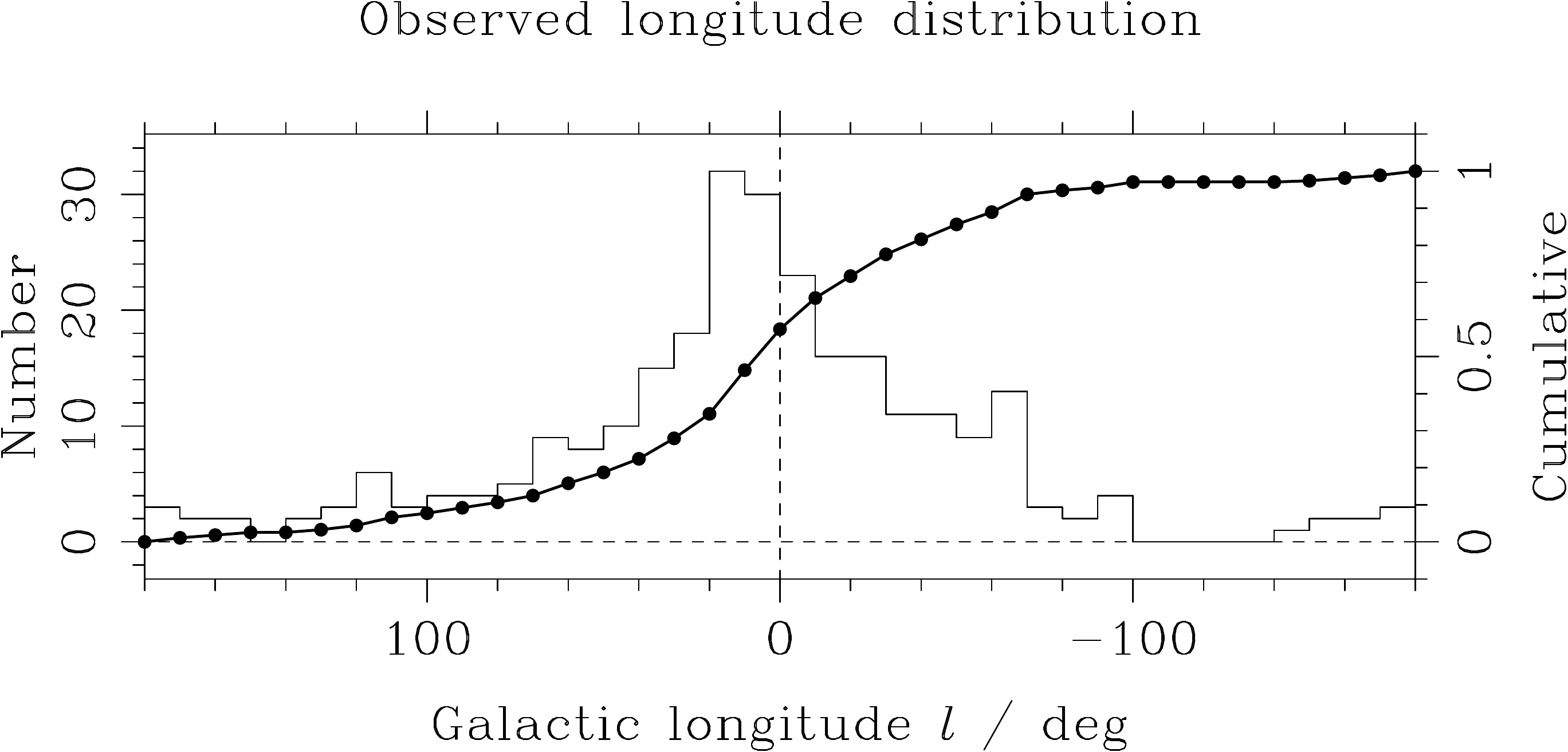}}
\leftline{\qquad b)}
\centerline{\includegraphics[viewport=54 0 378 788,clip=,
              angle=270,width=8.5cm]{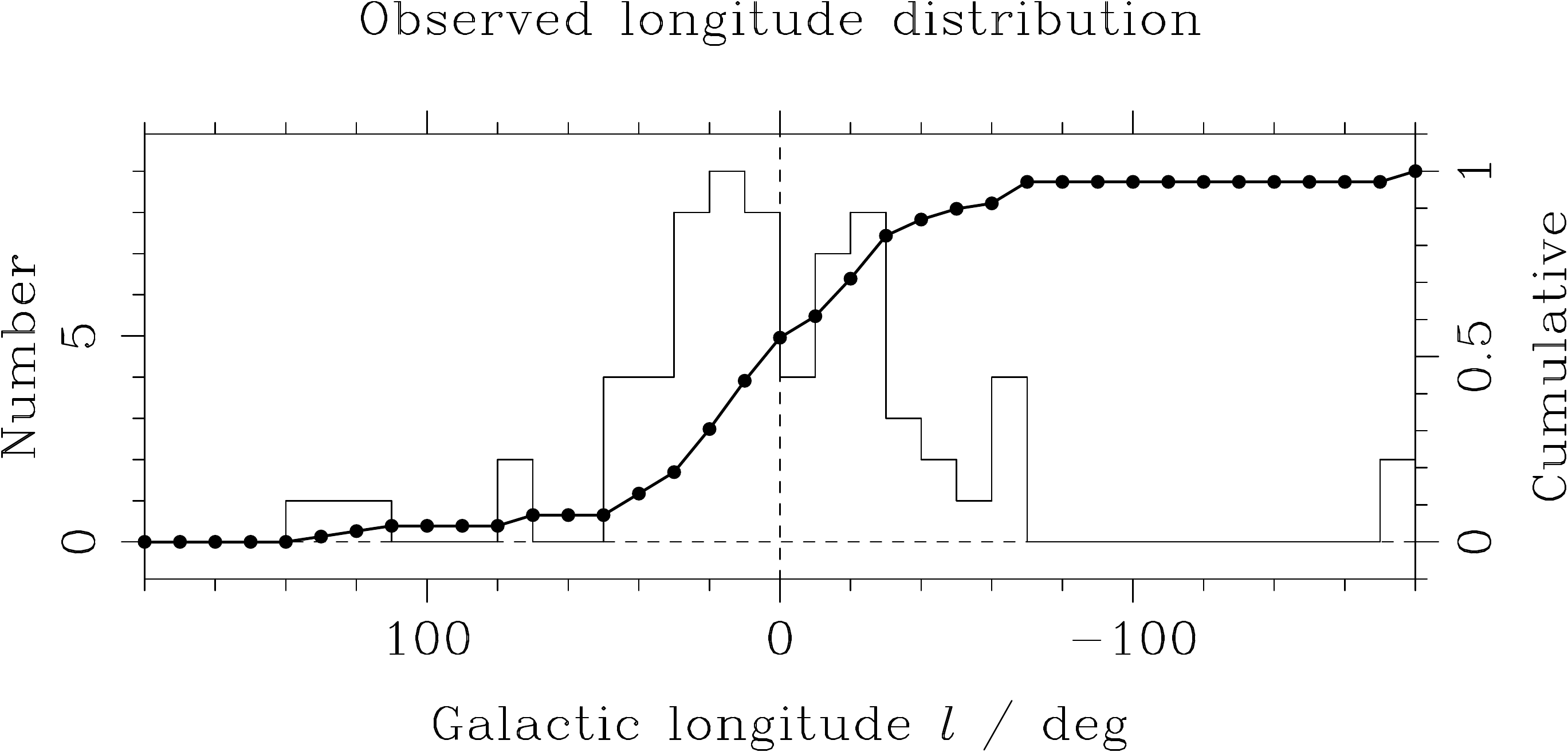}}
\caption{Histogram of the distribution of Galactic SNRs (left scale) and the
cumulative fraction (right scale) with Galactic longitude, for (a) all 294
catalogued SNRs, and (b) the 69 `bright' SNRs.\label{f:l-dist-o}}
\end{figure}

Figure~\ref{f:l-dist-o} shows the distribution of SNRs with Galactic longitude.
For all SNRs, Fig.~\ref{f:l-dist-o}(a) shows a clear asymmetry on different
sides of $l = 0^\circ$, with more SNRs identified in
the 1st and 2nd quadrants. This
is because (i) the 1st and 2nd quadrants are accessible to a large number of
northern hemisphere telescopes, and (ii) they include
the deep, multiwavelength survey of
$4\fdg5 < l < 22^\circ, |b| < 1\fdg25$ by \citet{2006ApJ...639L..25B}, as
discussed in Section~\ref{s:background}. The sample of 69 `bright' SNRs,
Fig.~\ref{f:l-dist-o}(b), shows a much smaller asymmetry close to $l=0^\circ$,
but this is not statistically significant (4 in the region $350^\circ \le l <
0^\circ$, and 8 -- including \SNR(0.0)+(0.0) in the region $0^\circ \le l <
10^\circ$). By Galactic quadrants, the numbers of bright SNRs are: 35 in the
1st, 3 in the 2nd, 2 in the 3rd, and 29 in the 4th. These, with Poisson errors,
do not show any asymmetry between either side of the Galactic Centre. This
implies that the nominal completeness surface brightness cutoff, although
derived from the Effelsberg 2.7-GHz -- which misses almost all of the 4th
quadrant -- is applicable there. Figure~\ref{f:l-dist-o} again shows that the
distribution of the `bright' SNRs is more localised towards $l = 0^\circ$ than
the distribution of all catalogued SNRs.

\subsection{The `$\Sigma{-}D$' relation}\label{s:Sigma-D}

Although distance determinations are available for some Galactic SNRs --
e.g.\ \citealt{2013ApJS..204....4P} give a compilation of distances to
60 SNRs from the literature -- the $\Sigma{-}D$ relation has been used
for some time (e.g.\ \citealt{1972A&A....18..169I, 1976MNRAS.174..267C})
to determine distances for other remnants. This is based on the fact
that, for remnants with known distances, the observed surface brightness
($\Sigma$) is larger for SNRs with smaller physical diameters ($D$).
This correlation is parameterised as
\begin{equation}
   \Sigma = A D^n,
\end{equation}
i.e.\ a straight line in the $\log{D}{-}\log{\Sigma}$ plane, with $A$
and $n$ determined from the properties of SNRs with known distances, The
\emph{observed} surface brightness, $\Sigma$ -- which is
distance-independent -- for a remnant without a distance determination
can then be used to determine its physical diameter, $D$. Hence via its
\emph{observed} angular size, $\theta$, its distance $d = D/\theta$ can
be determined. However -- as previously discussed
\citep{1991PASP..103..209G, 2005MmSAI..76..534G} -- there is a large
scatter in the observed $\Sigma{-}D$ distribution of SNRs, about an
order of magnitude in $D$ for a given $\Sigma$. In addition, small
and/or faint SNRs are more likely to have been missed in current
surveys, and be missing from the current Galactic SNR catalogue, due to
the observational selection effects discussed in \ref{s:selection}. So,
the true range of diameters for a given surface brightness may extend to
lower diameters. (On the other hand, for a given surface brightness the
upper limit to the range of diameters is not affected by selection
effects, at least down to the nominal surface brightness limit of
current catalogues.) Although the upper-right boundary of the observed
$\Sigma{-}D$ distribution of SNRs with known distances is useful to
provide an upper limit on the diameter of an individual SNR, distances
for individual SNRs derived from the $\Sigma{-}D$ relation are
imprecise.

Leaving aside issues with selection effects, if a $\Sigma{-}D$ relation is to
be used statistically to derive distances to individual SNRs from their
surface brightness, care has to be taken to use the appropriate form of
regression \citep{2005MmSAI..76..534G}. Since the $\Sigma{-}D$ relation is
used to derive $D$ from $\Sigma$ then a regression that minimises square
deviations in $\log{D}$ should be used, \emph{not} a regression than minimises
square deviations in $\log{\Sigma}$. For distributions with a large spread --
as is the case for SNRs with known distances -- different regressions give
significantly different fits (see \citealt{1990ApJ...364..104I,
1992ApJ...397...55F} and \citealt{2011ApJ...728...72F} for discussion of
different forms of least square regressions).

CB98 derived a $\Sigma{-}D$ relation with
\begin{equation}
  \Sigma \propto D^{-2.64 \pm 0.26},
\end{equation}
using 37 Galactic `shell' SNRs with known distances (or $\Sigma \propto D^{-2.38
\pm 0.26}$ if Cas~A (=\SNR(111.7)-(2.1)) is excluded). CB98 comment that these
$\Sigma{-}D$ slopes are significantly flatter than those derived in previous
studies, e.g.\ \citet{1979AuJPh..32...83M} obtained $\Sigma \propto D^{-3.8}$.
It is evident, however, that CB98 minimised square deviations in $\log{\Sigma}$,
whereas \citet{1979AuJPh..32...83M} minimised square deviations in $\log{D}$, so
these $\Sigma{-}D$ slopes are not directly comparable. Re-fitting the SNRs with
known distances used by CB98, minimising deviations in $\log{D}$ rather than
$\log{\Sigma}$ produces much steeper $\Sigma{-}D$ slopes, which \emph{are} in
good agreement with earlier results (e.g.\ those of
\citeauthor{1979AuJPh..32...83M}). For all 37 `shell' SNRs
\begin{equation}
  \Sigma \propto D^{-3.58 \pm 0.33},
\end{equation}
(or $\Sigma \propto D^{-3.37 \pm 0.35}$ if Cas~A is excluded). This
difference in the slope of the derived $\Sigma{-}D$ relation depending
on the form of regression used has an important consequence for the
Galactic SNR distribution derived by CB98. There is a systematic bias in
the diameters and hence distances derived for SNRs, with
fainter/brighter SNRs having distances that are too large/small
respectively. Since there are more faint SNRs than bright SNRs, then the
derived distribution of SNRs in the Galaxy will be systematically spread
out too. In $\Sigma{-}D$ studies in the literature it is not always
clear what form of regression was used. Re-analysing the published lists
of SNRs with known distances implies that others (e.g.\
\citealt{1981A&A....93...43G, 1985ApJ...295L..13H, 2005MNRAS.360...76A,
2007Ap&SS.307..423S}) have\footnote{This is contrary to the statement in
\citet{2013ApJS..204....4P} that all previous $\Sigma{-}D$ studies used
regression minimising deviations in $\log\Sigma$.}, like CB98, used
regressions which minimise deviations on $\log{\Sigma}$ rather than
$\log{D}$.

\citet{2013ApJS..204....4P} discussed the fitting of a $\Sigma{-}D$
relation to Galactic SNRs with known distances, and concluded from Monte
Carlo simulations that `orthogonal regression' gives the best result.
This conclusion is to be expected, given that in their Monte Carlo
simulations \citeauthor{2013ApJS..204....4P} assumes errors in $\log{D}$
and $\log{\Sigma}$ of same magnitude, which is not realistic. In
practice, the errors in surface brightness -- which depends on the
uncertainties in flux density and angular size -- are likely to be
smaller than the errors in distances (as
\citeauthor{2013ApJS..204....4P} say in their introduction). When the
errors in $\log{D}$ and $\log{\Sigma}$ are different, then the
orthogonal fitting does not produce the correct result. However, if the
$\Sigma{-}D$ relation is to be used to derive distances to individual
SNRs, then a regression minimising deviations in $\log{D}$ should be
used to obtain the best result.

\begin{figure}
\centerline{\includegraphics[viewport=54 0 378 788,clip=,
              angle=270,width=8.5cm]{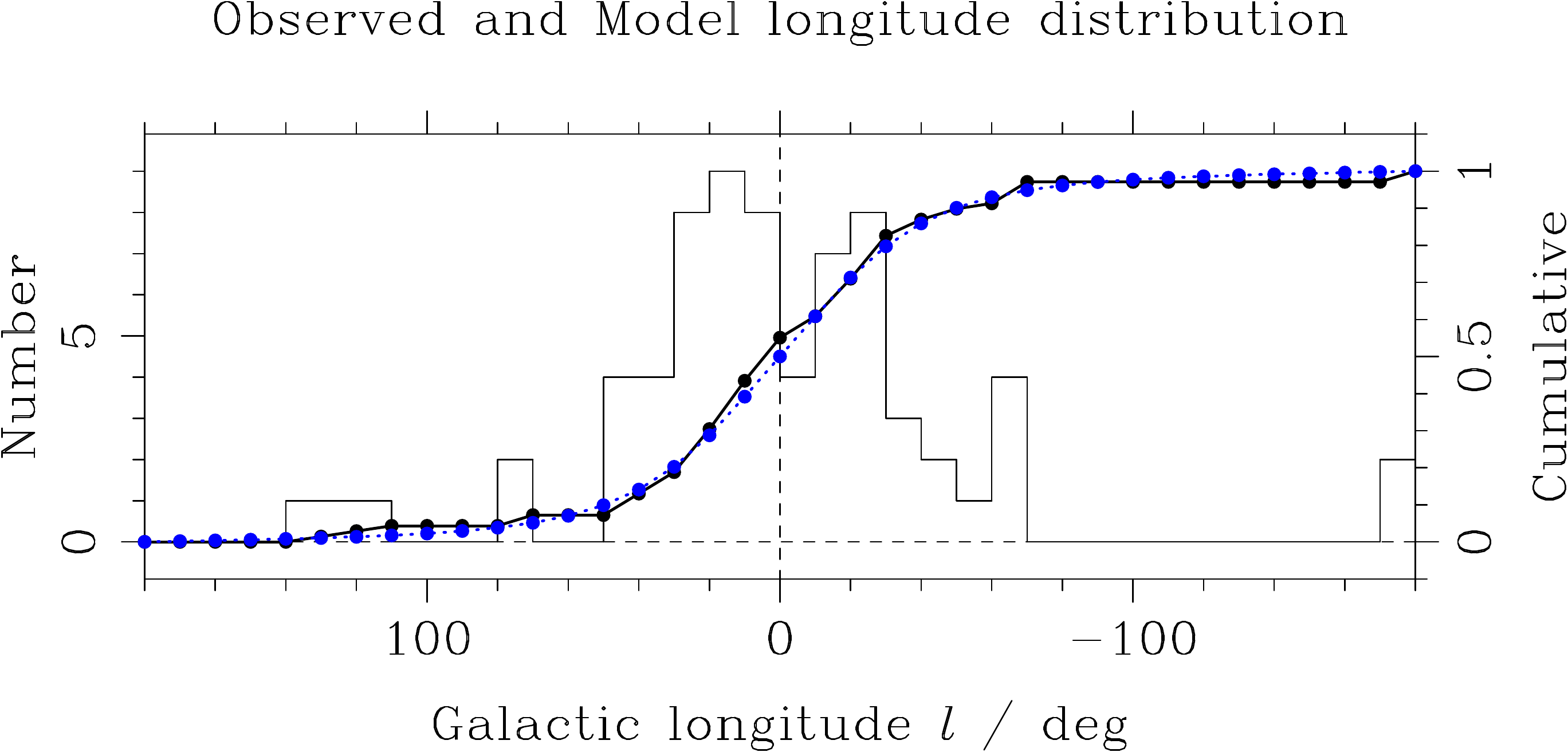}}
\caption{Histogram of the distribution of Galactic SNRs (left scale) and the
cumulative fraction -- solid black line -- (right scale) with Galactic longitude
for the 69 `bright' SNRs. The dashed blue line is the cumulative fraction for
the best-fitting power-low/exponential model (with $\alpha = 1.09$ and
$\beta = 3.87$).\label{f:sd-best}}
\end{figure}

\section{The Galactic radial distribution of SNRs}\label{s:radial}

Given the limitations of the $\Sigma{-}D$ relation and the selection
effects that apply to the identification of SNRs, rather than deriving
the distribution of SNRs in the Galaxy directly -- as done by CB98 -- an
alternative approach is to consider the $l$-distribution of sample SNRs.
This can be compared with the expected $l$-distribution from various
models.

Here I choose the sample of 69 `bright' SNRs, above the nominal surface
brightness limit of $\Sigma_{\rm 1~GHz} = 10^{-20}~\SigmaUnit$, where
the current catalogue of Galactic SNRs is thought to be nearly complete.
This method has the advantage of avoiding all of the uncertainties from
the $\Sigma{-}D$ relation, but has the disadvantage of only using a
fraction of the total number of catalogued Galactic SNRs. I include
checks on how dependent the results are on a choice of surface
brightness limit, or exclusion of regions close to $l=0^\circ$. As noted
above, a single surface brightness limit is not appropriate at all
Galactic latitudes, and some SNRs above this nominal limit may well
be missed in the brightest part of the Galactic plane, i.e.\ at
latitudes near $l=0^\circ$. Moreover, the difficulty of identifying
small angular size remnants is also more of an issue nearer $l=0^\circ$.
Nevertheless, this sample is not as strongly affected by selection effects
as the complete catalogue.

\begin{figure}
%
%
%
\centerline{\includegraphics[viewport=54 0 378 788,clip=,
              angle=270,width=8.5cm]{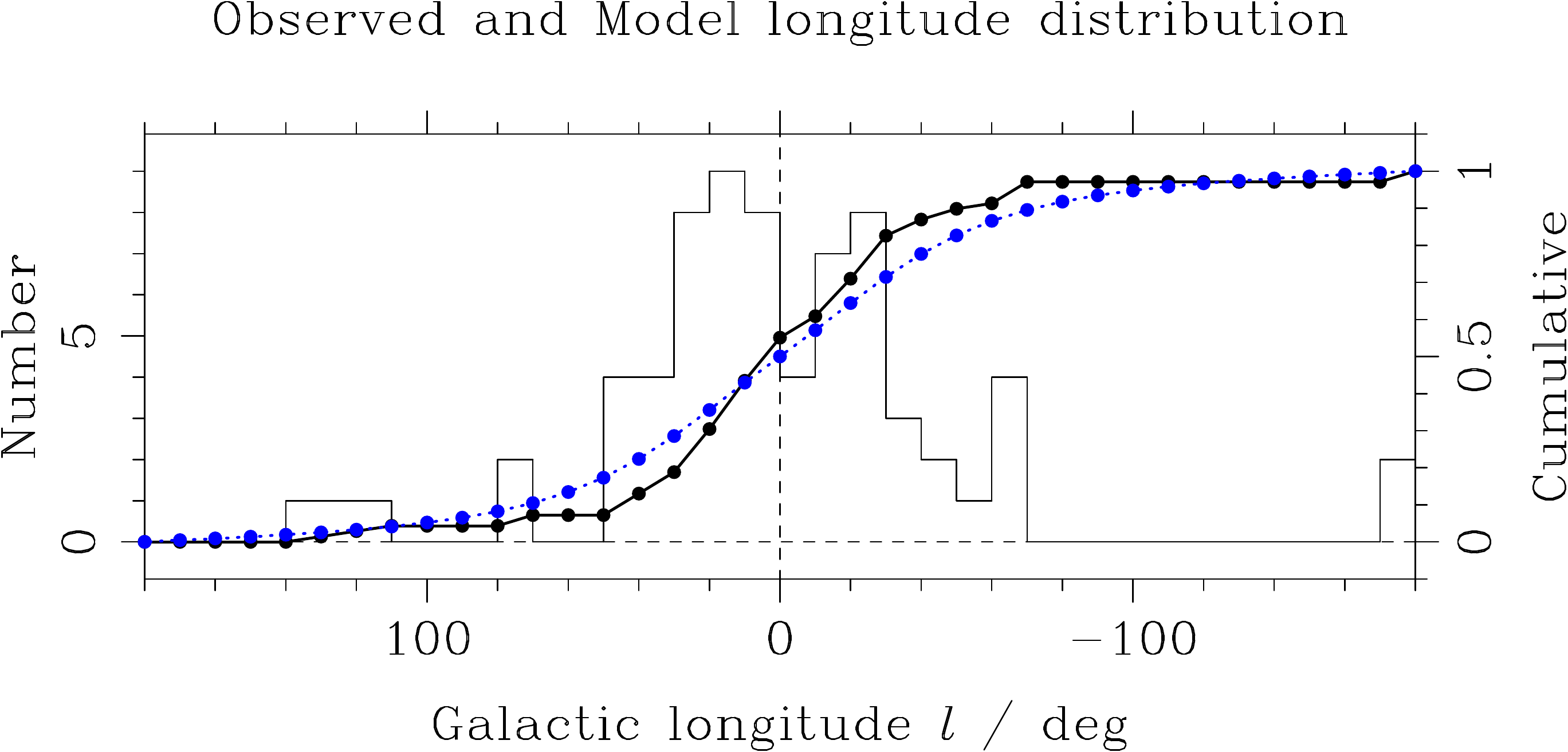}}
\caption{As Fig.~\ref{f:sd-best}, but the dashed blue line is the
cumulative fraction for the best-fitting power-low/exponential model from
CB98 (with $\alpha = 2.00$ and $\beta = 3.53$).\label{f:sd-CB98}}
\end{figure}

\begin{figure}
\centerline{\includegraphics[width=8.5cm]{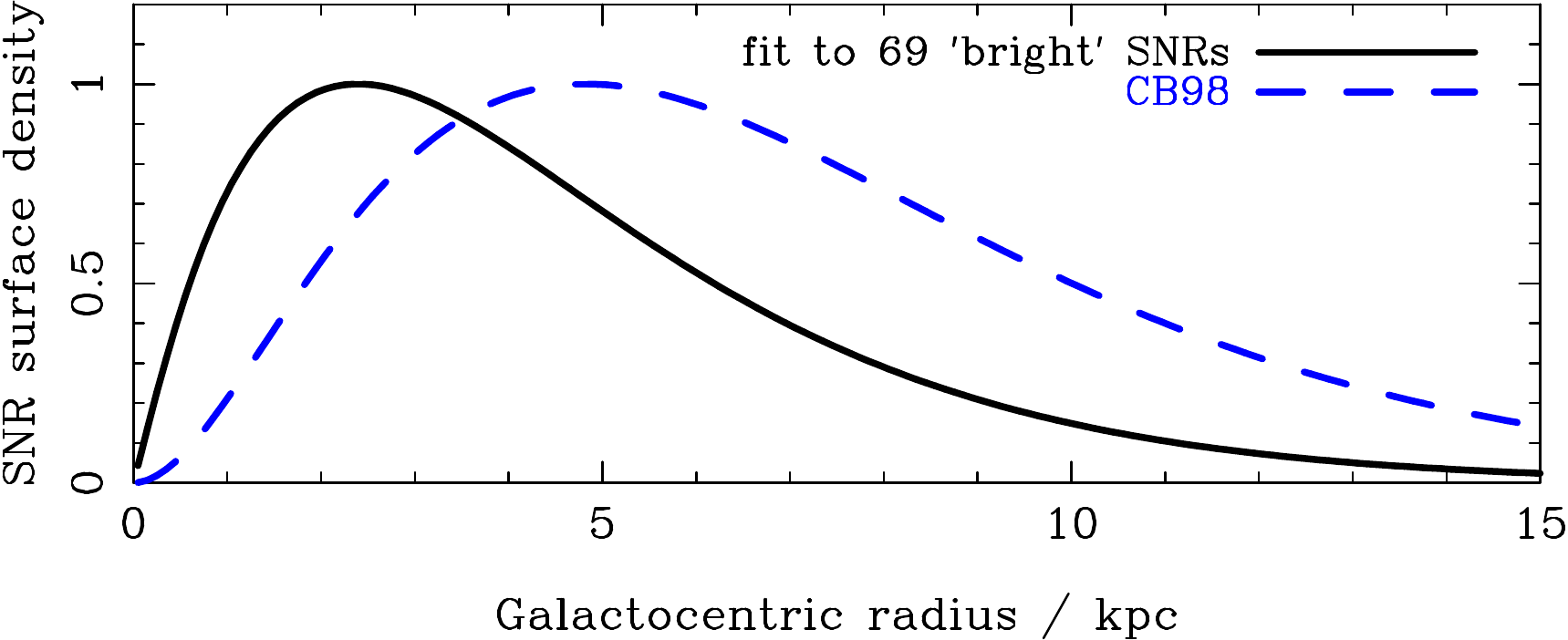}}
\caption{Normalised surface density of SNRs with Galactocentric radius
for power-low/exponential models: (i) solid line, the best fit to
$l$-distribution of 69 bright SNRs, and (ii) dashed line, from
CB98.}\label{f:radial-best}
\end{figure}

Here I use a power law/exponential model for the Galactic surface
density of SNRs with Galactocentric radius, $R$, with
\begin{equation}
  \propto \left( \frac{R}{R_\odot} \right)^\alpha
          \exp \left( -\beta \frac{(R-R_\odot)}{R_\odot} \right),
\end{equation}
where $R_\odot=8.5$~kpc
(i.e.\ a cylindrically symmetrical distribution about the Galactic
Centre). Given the limited number of SNRs in the sample of `bright'
remnants, no attempt is made here to constrain the form of the
distribution perpendicular to the Galactic plane. This model for the
surface density of SNRs tends to a zero density towards the Galactic
Centre, which better matches the distributions derived for pulsars and
star formation in the Galaxy (e.g.\ \citealt{1994MNRAS.268..595J,
2000A&A...358..521B, 2004Ap&SS.289..363P, 2004A&A...422..545Y}) than,
for example, a simple Gaussian distribution (which I used in
\citealt{1996ssr..conf..341G}). This power law/exponential is one of the
two models used by CB98, who obtained best-fitting parameters of
$\alpha=2.00 \pm 0.67$, $\beta = 3.53 \pm 0.77$.

\begin{table*}
\caption{Best-fitting power-law/exponential models for different samples of
Galactic SNRs, and other power-law/exponential models for
comparison.}\label{t:fits}
\begin{tabular}{lcccccc}\toprule
                      &  $\Sigma_{\rm 1~GHz}$ cut-off & $l$-range & number  & \multicolumn{2}{c}{fit parameters} & proportion inside \\\cmidrule{5-6}
                      &    / $\SigmaUnit$             &           & of SNRs & $\alpha$ &  $\beta$                &    Solar Circle   \\\midrule
bright sample         &       $10^{-20}$      &        all        &    69   &  1.09  & 3.87  & 73\% \\
brighter sample       & $2 \times 10^{-20}$   &        all        &    44   &  0.51  & 2.91  & 70\% \\
fainter sample        & $5 \times 10^{-21}$   &        all        &   103   &  1.49  & 4.60  & 77\% \\
omit near GC sample   &     $10^{-20}$        & $|l| > 10^\circ$  &    57   &  0.0   & 2.76  & 77\% \\\midrule
CB98 best fit         &     $10^{-20}$        &        all        &    69   &  2.00  & 3.53  & 49\% \\
CB98 $\alpha$ fixed   &     $10^{-20}$        &        all        &    69   &  2.00  & 5.11  & 76\% \\
CB98 $\beta$  fixed   &     $10^{-20}$        &        all        &    69   &  0.85  & 3.53  & 73\% \\\bottomrule
\end{tabular}\\[5pt]
\begin{minipage}{12cm}
\small Notes: (a) for the first four rows, the $\alpha$ and $\beta$
values are for the best fit; (b) the fifth row uses the $\alpha$ and
$\beta$ best-fitting values from CB98; (c) the final two entries use one
or other of the best0fitting values for $\alpha$ and $\beta$ from CB98
and varies the other for a best fit.
\end{minipage}
\end{table*}

The best-fitting power-law/exponential model to the 69 bright remnants,
is shown in Fig.~\ref{f:sd-best}, which has $\alpha = 1.09$ and $\beta =
3.87$.
The statistic used for the fitting was the least sum of squares
of the differences between the observed and model cumulative fractions,
summed over all latitude bins, i.e.\ if the observed and model
cumulative fractions are $f_{\rm o}(i)$ and $f_{\rm m}(i)$ respectively,
for the $i$th bin, the minimum of $\sum_{i} (f_{\rm o} - f_{\rm m})^2$.
(The fitting was made with a model distribution of one million SNRs, out
to a Galactocentric radius of 50~kpc, using bins of $10^\circ$ in $l$.)
The sum of squares misfit for the $\alpha = 1.09$ and $\beta =
3.87$ bets fitting model is $0.0127$. Figure~\ref{f:sd-CB98} shows the
power-law/exponential model derived by CB98, which is a poor fit to the
observed distribution of the 69 bright SNRs, with a sum of squares
misfit of 0.0841. This clearly represents a broader distribution (e.g.\
a flatter slope of the cumulative distribution near $l=0^\circ$ than
what is observed for the `bright' remnants) than that shown in
Fig.~\ref{f:sd-best}. This is not surprising given the systematic bias
in the distances derived from the $\Sigma{-}D$ relation used by CB98
which was noted in Section~\ref{s:Sigma-D}. The parameters for
the best-fitting power-law/exponential model for the sample of 69 bright
SNRs, and that from CB98 are shown in Table~\ref{t:fits}. The
broader CB98 distribution is also evident in Fig.~\ref{f:radial-best},
which shows the surface density distribution of SNRs with Galactocentric
radius for CB98's model, and the best fit to the 69 bright remnants. For
CB98's best-fitting model, only 49~per~cent of SNRs are inside the Solar
Circle, compared with 73~per~cent for the best fit to the 69 bright
remnants.

The parameters of the best-fitting model are not well defined, as there is a
strong degeneracy between the parameters, e.g.\ it is possible to obtain
almost as good fits with, $\alpha = 2.00$, $\beta = 5.11$ or $\alpha =
0.85, \beta = 3.53$ (i.e.\ fixing one parameter to the value obtained by
CB98, and varying the other), which have sum of squares misfits of
0.0135 and 0.0128 respectively. The distribution with $\alpha = 2.00$,
$\beta = 5.10$ peaks at a slightly larger radius from GC than the
best-fitting distribution with $\alpha = 1.10$ and $\beta = 3.90$, but has a
similar fraction of SNRs, 76~per~cent, inside the Solar Circle.
Conversely the distribution with $\alpha = 2.00$, $\beta = 5.10$ peaks
closer to the GC, but again has a similar fraction of SNRs, 73~per~cent,
inside the Solar Circle.

Also, the best-fitting model depends on (i) the accuracy of the value of
the surface brightness cut-off chosen to define the sample of `bright'
SNRs, and also (ii) the assumption that the remaining selection effects,
which are more important close to $l = 0^\circ$ are not important. To
investigate these I have done the analysis with different $\Sigma_{\rm
1~GHz}$ cut-off values (lower and higher by a factor of two), and
excluding SNRs within $10^\circ$ of $l = 0^\circ$. Varying the surface
brightness by a factor of two -- the `brighter' and `fainter'
samples in Table~\ref{t:fits} -- does give rather different parameters,
and hence different radial distribution, but the fraction of SNRs within
the Solar Circle does not change strongly. When the region $|l|
\le 10^\circ$ is excluded -- the `omit near GC' sample in
Table~\ref{t:fits} -- the best-fitting model is a pure exponential (but
note that this fitting does not depend on $R \la 1.5$~kpc). This
suggests there are indeed residual selection effects near $l=0^\circ$.
Nevertheless, the main result, that the distribution obtained by CB98 is
too broad compared with bright SNRs still holds. The conclusion is
strengthened by the fact that any remaining incompleteness of the sample
of 69 bright SNRs will be concentrate towards $l=0^\circ$.

Finally, it should be noted that the constraints presented here apply to
the distribution of observed SNRs. If the observability of SNRs above
the radio surface brightness limit depends on a property -- e.g.\
ambient density -- which varies with Galactocentric radius, then this
will mean the distribution of SNRs is not the same as their parent SNe.

\section{Conclusions}\label{s:conclusions}

Here I have discussed some of the properties of the most recent
catalogue of Galactic SNRs, which contains 294, particularly (i) the
selection effects that apply to the completeness of the catalogue, and
(ii) issues with using the `$\Sigma{-}D$' relation to derive distances
to individual remnants. By comparison of the distribution in Galactic
longitude of a sample of 69 `bright' SNRs -- which are not strongly
affected by selection effects -- with that expected from models,
constraints are placed on the distribution of the SNRs with
galactocentric radius (using a power-law/exponential model). It is shown
that the widely used distribution derived by \citet{1998ApJ...504..761C}
is too broad.

\section*{Acknowledgements}

I thank various colleagues for useful comments on this work. I am
grateful to the National Centre for Radio Astrophysics of the Tata
Institute of Fundamental Research, Pune for their hospitality during a
recent visit, during which the work for this paper was completed.

%
%

\bsp

\label{lastpage}

\begin{thebibliography}{}

\bibitem[Arbutina \& Uro{\v s}evi{\'c}(2005)Arbutina \& Uro{\v s}evi{\'c}]{2005MNRAS.360...76A}
  Arbutina B., Uro{\v s}evi{\'c} D.,
    2005, MNRAS, 360, 76

\bibitem[Bell(2014)Bell]{2014BrJPh..44..415B}
  Bell A.~R.,
    2014, Braz. J.\ Phys., 44, 415

\bibitem[Bocchino et al.(2010)Bocchino, Bandiera \& Gelfand]{2010A&A...520A..71B}
  Bocchino F., Bandiera R., Gelfand J.,
    2010, A\&A, 520, A71

\bibitem[Borkowski et al.(2014)Borkowski et al.]{2014ApJ...790L..18B}
  Borkowski K.~J., Reynolds S.~P., Green D.~A., Hwang U., Petre R.,
  Krishnamurthy K., Willett R.,
    2014, ApJ, 790, L18

\bibitem[Brogan et al.(2006)Brogan et al.]{2006ApJ...639L..25B}
  Brogan C.~L., Gelfand J.~D., Gaensler B.~M., Kassim N.~E., Lazio T.~J.~W.,
    2006, ApJ, 639, L25

\bibitem[Bronfman et al.(2000)Bronfman et al.]{2000A&A...358..521B}
  Bronfman L., Casassus S., May J., Nyman L.-{\AA}.,
    2000, A\&A, 358, 521

\bibitem[Cameron \& Kulkarni(2007)Cameron \& Kulkarni]{2007ApJ...665L.135C}
  Cameron P.~B., Kulkarni S.~R.,
    2007, ApJ, 665, L135

\bibitem[Case \& Bhattacharya(1996)Case \& Bhattacharya]{1996A&AS..120C.437C}
  Case G., Bhattacharya D.,
    1996, A\&AS, 120, 437

\bibitem[Calore et al.(2015)Calore, Cholis \& Weniger]{2015JCAP...03..038C}
  Calore F., Cholis I., Weniger C.,
    2015, J. Cosmology Astropart. Phys., 3, 038

\bibitem[Case \& Bhattacharya(1998)Case \& Bhattacharya]{1998ApJ...504..761C}
  Case G.~L., Bhattacharya D.,
    1998, ApJ, 504, 761

\bibitem[Clark \& Caswell(1976)Clark \& Caswell]{1976MNRAS.174..267C}
  Clark D.~H., Caswell J.~L.,
    1976, MNRAS, 174, 267

\bibitem[Feigelson \& Babu(1992)Feigelson \& Babu]{1992ApJ...397...55F}
  Feigelson E.~D., Babu G.~J.,
    1992, ApJ, 397, 55

\bibitem[Feigelson \& Babu(2011)Feigelson \& Babu]{2011ApJ...728...72F}
  Feigelson E.~D., Babu G.~J.,
    2011, ApJ, 728, 72

\bibitem[F\"urst et al.(1990)F\"urst et al.]{1990A&AS...85..691F}
  F\"urst E., Reich W., Reich P., Reif K.,
    1990, A\&AS, 85, 691

\bibitem[G\"obel et al.(1981)G\"obel, Hirth, \& F\"urst]{1981A&A....93...43G}
  G\"obel W., Hirth W., F\"urst E.,
    1981, A\&A, 93, 43

\bibitem[Green(1984)Green]{1984MNRAS.209..449G}
  Green D.~A.,
    1984, MNRAS, 209, 449

\bibitem[Green(1985)Green]{1985MNRAS.216..691G}
  Green D.~A.,
    1985, MNRAS, 216, 691

\bibitem[Green(1986)Green]{1986MNRAS.219P..39G}
  Green D.~A.,
    1986, MNRAS, 219, 39P

\bibitem[Green(1988)Green]{1988Ap&SS.148....3G}
  Green D.~A.,
    1988, Ap\&SS, 148, 3

\bibitem[Green(1991)Green]{1991PASP..103..209G}
  Green D.~A.,
    1991, PASP, 103, 209

\bibitem[Green(1996a)Green]{1996ssr..conf..419G}
  Green D.~A.,
    1996a, in McCray R., Wang Z., eds, IAU Colloquium 145, Supernovae and
    supernova remnants, CUP, p.~419

\bibitem[Green(1996a)Green]{1996ssr..conf..341G}
  Green D.~A.,
    1996b, in McCray R., Wang Z., eds, IAU Colloquium 145, Supernovae and
    supernova remnants, CUP, p.~341

\bibitem[Green(2004)Green]{2004BASI...32..335G}
  Green D.~A.,
    2004, Bull. Astron. Soc. India, 32, 335

\bibitem[Green(2005)Green]{2005MmSAI..76..534G}
  Green D.~A.,
    2005, Mem. Soc. Astron. Ital., 76, 534

\bibitem[Green(2009)Green]{2009BASI...37...45G}
  Green D.~A.,
    2009, Bull. Astron. Soc. India, 37, 45

\bibitem[Green(2012)Green]{2012AIPC.1505....5G}
  Green D.~A.,
    2012, in Aharonian F.~A., Hofmann W., Rieger F.~M., eds, AIP Conference
    Proceedings, Volume 1505, High energy gamma-ray astronomy, AIPC, p.~5

\bibitem[Green(2014a)Green]{2014BASI...42...47G}
  Green D.~A.,
    2014a, Bull. Astron. Soc. India, 42, 47

\bibitem[Green(2014b)Green]{2014IAUS..296..188G}
  Green D.~A.,
    2014b, in Ray A., McCray R.~A., eds, IAU Symposium 296, Supernova
    Environmental Impacts, CUP, p.~188

\bibitem[Green et al.(1999)Green et al.]{1999ApJS..122..207G}
  Green A.~J., Cram L.~E., Large M.~I., Ye T.,
    1999, ApJS, 122, 207

\bibitem[Green et al.(2008)Green et al.]{2008MNRAS.387L..54G}
  Green D.~A., Reynolds S.~P., Borkowski K.~J., Hwang U., Harrus I., Petre R.,
    2008, MNRAS, 387, L54

\bibitem[Huang \& Thaddeus(1985)Huang \& Thaddeus]{1985ApJ...295L..13H}
  Huang Y.-L., Thaddeus P.,
    1985, ApJ, 295, L13

\bibitem[Ilovaisky \& Lequeux(1972)Ilovaisky \& Lequeux]{1972A&A....18..169I}
  Ilovaisky S.~A., Lequeux J.,
    1972, A\&A, 18, 169

\bibitem[Isobe et al.(1990)Isobe et al.]{1990ApJ...364..104I}
  Isobe T., Feigelson E.~D., Akritas M.~G., Babu G.~J.,
    1990, ApJ, 364, 104

\bibitem[Johnston(1994)Johnston]{1994MNRAS.268..595J}
  Johnston S.,
    1994, MNRAS, 268, 595

\bibitem[Kumar \& Eichler(2014)Kumar \& Eichler]{2014ApJ...785..129K}
  Kumar R., Eichler D.,
    2014, ApJ, 785, 129

\bibitem[Lee et al.(2011)Lee et al.]{2011APh....35..211L}
  Lee S.-H., et al.,
    2011, Astropart. Phys, 35, 211

\bibitem[Li et al.(2011)Li et al.]{2011MNRAS.412.1473L}
  Li W., Chornock R., Leaman J., Filippenko A.~V., Poznanski D., Wang X., Ganeshalingam M., Mannucci F.,
    2011, MNRAS, 412, 1473

\bibitem[Milne(1979)Milne]{1979AuJPh..32...83M}
  Milne D.~K.,
    1979, Aust. J. Phys., 32, 83

\bibitem[Onello et al.(1995)Onello et al.]{1995ApJ...449L.127O}
  Onello J.~S., Depree C.~G., Phillips J.~A., Goss W.~M.,
    1995, ApJ, 449, L127

\bibitem[Paladini et al.(2004)Paladini, Davies, \& De Zotti]{2004Ap&SS.289..363P}
  Paladini R., Davies R., De Zotti G.,
    2004, Ap\&SS, 289, 363

\bibitem[Pavlovi{\'c} et al.(2013)Pavlovi{\'c} et al.]{2013ApJS..204....4P}
  Pavlovi{\'c} M.~Z., Uro{\v s}evi{\'c} D., Vukoti{\'c} B., Arbutina B.,
  G{\"o}ker {\"U}.~D.,
    2013, ApJS, 204, 4

\bibitem[Reich et al.(1985)Reich et al.]{1985A&A...151L..10R}
  Reich W., F\"urst E., Altenhoff W.~J., Reich P., Junkes N.,
    1985, A\&A, 151, L10

\bibitem[Reich et al.(1988)Reich et al.]{1988srim.conf..293R}
  Reich W., F{\"u}rst E., Reich P., Junkes N.,
    1988, in Roger R.~S., Landecker T.~L., eds, IAU Colloquium 101,
    Supernova Remnants and the Interstellar Medium, CUP, p.~293

\bibitem[Reich et al.(1990)Reich et al.]{1990A&AS...85..633R}
  Reich W., F\"urst E., Reich P., Reif K.,
    1990, A\&AS, 85, 633

\bibitem[Reynolds et al.(2008)Reynolds et al.]{2008ApJ...680L..41R}
  Reynolds S.~P., Borkowski K.~J., Green D.~A., Hwang U., Harrus I.,
  Petre R.,
    2008, ApJ, 680, L41

\bibitem[Sabin et al.(2013)Sabin et al.]{2013MNRAS.431..279S}
  Sabin L., et al.,
    2013, MNRAS, 431, 279

\bibitem[Stupar et al.(2007)Stupar et al.]{2007Ap&SS.307..423S}
  Stupar M., Filipovi{\'c} M.~D., Parker Q.~A., White G.~L., Pannuti T.~G.,
  Jones P.~A.,
    2007, Ap\&SS, 307, 423

\bibitem[Vladimirov et al.(2012)Vladimirov et al.]{2012ApJ...752...68V}
  Vladimirov A.~E., J{\'o}hannesson G., Moskalenko I.~V., Porter T.~A.,
    2012, ApJ, 752, 68

\bibitem[Yusifov \& K{\"u}{\c c}{\"u}k(2004)Yusifov \& K{\"u}{\c c}{\"u}k]{2004A&A...422..545Y}
  Yusifov I., K{\"u}{\c c}{\"u}k I.,
    2004, A\&A, 422, 545

\bibitem[Xu et al.(2013)Xu et al.]{2013A&A...559A..81X}
  Xu W.~F., Gao X.~Y., Han J.~L., Liu F.~S.,
    2013, A\&A, 559, A81

\end{thebibliography}
\end{document}